\documentclass[12pt]{iopart}
\usepackage{graphicx}
\newcommand\beq{\begin{equation}}
\newcommand\eeq{\end{equation}}
\newcommand\bea{\begin{eqnarray}}
\newcommand\eea{\end{eqnarray}}

\begin{document}
\title {Cold atoms at unitarity and inverse square interaction}
\author{R.K. Bhaduri$^{1,2}$, M.V.N. Murthy $^{2}$ and M.K. 
Srivastava $^3$}
\address{$^1$ Department of Physics and Astronomy, McMaster 
University,
Hamilton L8S 4M1, Canada}
\address{$^2$ The Institute of Mathematical Sciences, Chennai 600113, 
India}
\address{$^3$ Institute Instrumentation Centre, Indian Institute of 
technology, Roorkee 247667, India}
\date{\today}

\begin{abstract}

Consider two identical atoms in a spherical harmonic oscillator 
interacting with a zero-range interaction which is tuned to produce an 
s-wave zero-energy bound state. The quantum spectrum of the system is 
known to be exactly solvable. We note that the same partial wave quantum 
spectrum is obtained by the one-dimensional scale-invariant inverse square 
potential. Long known as the Calogero-Sutherland-Moser (CSM) model, it 
leads to Fractional Exclusion Statistics (FES) of Haldane and Wu. The 
statistical parameter is deduced from the analytically calculated second 
virial coefficient. When FES is applied to a Fermi gas at unitarity, it 
gives good agreement with experimental data without the use of any free 
parameter.

\end{abstract}
\pacs{~03.75.Ss, 05.30.-d}
\submitto{\JPB}
\maketitle

\section{Introduction}

Experimental investigations on ultra-cold gas of fermionic atoms near 
Feshbach resonance, in recent years, have opened new avenues to address 
and understand the problems in strongly correlated fermionic 
systems~\cite{regal}. Two identical fermionic atoms trapped in different
hyperfine states may still interact in the relative s-state. 
Low energy properties of such a gas at low density 
are determined by the scattering length $a$, the number density $n$ and the 
temperature $T$. The effective attraction between the atoms near a Feshbach 
resonance may be increased continuously by varying an applied magnetic 
field to reduce the Zeeman splitting between the states occupied by the 
atoms and the resonance. The scattering length $a$ goes from a small 
negative to a positive value. The unitary limit is achieved when $|a|$ is 
infinite at the transition when $a$ changes its sign resulting in a 
zero-energy two-body bound state. This defines the unitary limit and the 
behaviour is expected to be universal (scale independent) in this 
limit~\cite{baker}.

\begin{figure}[htp] 
\centering 
\includegraphics[width=6cm]{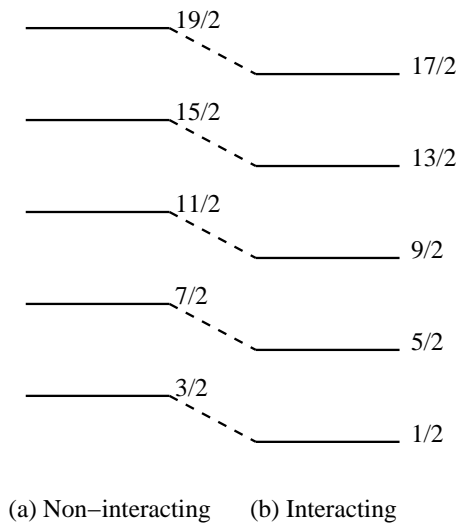} 
\caption{The s-wave spectrum of few lowest states of two identical 
fermionic atoms are shown: (a) Spectrum of non-interacting fermions. (b) 
spectrum of interacting fermions at unitarity. Note the spectrum is simply 
shifted down by one unit of energy. The energies are given in units of 
$\hbar\omega$, and the spacing between two-adjacent s-wave states is 
$2 \hbar\omega$.} 
\label{fig1} 
\end{figure}

Recently, Liu {\it{et al}}~\cite{liu} have calculated the virial expansion 
coefficients of the equation of state of a strongly correlated trapped Fermi 
gas on either side of the unitary limit, extending the work of Ho and 
Mueller~\cite{ho}. The latter had earlier developed a virial expansion 
up to the 
second virial coefficient to study the universal behaviour of a homogeneous 
gas at unitarity. The central point of these investigations, for our 
purpose, is the fact that the second and third virial coefficients, when 
plotted as a function of the interaction parameter (scattering length in 
this case), become temperature independent at unitarity, testifying to the 
universal nature of the quantum gas at this limit. More over, the two-body 
bound state spectrum (in the s-state) for harmonic confinement at 
unitarity, shown in Fig.(\ref{fig1}), exhibits the striking property of an 
overall shift in the energy levels due to the interaction. 
This is a hallmark of the inverse square interaction, which, in 
one-dimension, also leads to fractional exclusion statistics(FES ) as 
defined by Haldane~\cite{haldane}. In FES, the occupancy factor $n_i(T)$ of a
single-particle state with energy $\epsilon_i$ at temperature $T$ is 
given by~\cite{wu}
\beq
n_i=(w_i+g)^{-1}, 
\label{late}
\eeq
where the distribution function $w$ satisfies the nonlinear relation 
\beq
w_i^g (1+w_i)^{1-g}=\exp[(\epsilon_i-\mu)\beta]~.
\label{late2} 
\eeq
In the above, $g\geq 0$ is the (temperature-independent) statistical parameter, 
$\beta=1/(k_B T)$, and $\mu$ the chemical potential.
The parameter $g$ is based on the 
rate at which the number of available states in a system of fixed size 
decreases as more and more particles are added to it. As such,    
$g$ assumes values 0 and 1 for bosons and fermions respectively, 
because the addition of one particle reduces the number of available 
states by $g$~\cite{nayak}.

As may be deduced from Eqs. (\ref{late}, \ref{late2}) the  occupancy 
factor $n_i$ at $T=0$ for an ideal
FES gas is specially simple, and is given by $n_i=1/g$ up to 
$k_i\leq \tilde{k}_F$, and zero
otherwise, where $\tilde{k}_F$ is the shifted Fermi wave number. 
The relationship between $\tilde{k}_F$ and the fermionic $k_F$ is
obtained by noting that the particle number $N$ is
\beq 
N=\frac{1}{g}\int_0^{\tilde{k}_F}~4\pi k^2
dk=\frac{4\pi}{3}\frac{1}{g} \tilde{k}_F^3~.
\eeq
But for fermions $N=\frac{4\pi}{3} k_F^3$. For a fixed $N$, we thus get
$\tilde{k}_F=g^{1/3} k_F$. A similar calculation for the energy $E$ of
an ideal FES gas gives $E=\frac{1}{g}\frac{4\pi}{5} 
\frac {\tilde{k}_F^5}{2M}$~. Eliminating $\tilde{k}_F$, we then get   
\beq
\frac{E}{N} =\xi\frac{3 \hbar^2 k_F^2}{10 M},
\label{eq1}
\eeq
where $\xi=g^{2/3}$.

In an earlier investigation~\cite{bhaduri}, the energy per particle and 
the chemical potential at finite temperature of the quantum Fermi gas at 
unitarity were calculated by mapping the interacting fermionic system to a 
system of non-interacting quasi-particles obeying FES. In a subsequent
paper~\cite{bhaduri2}, some properties of few-body systems were
calculated in the same scheme. The 
statistical parameter $g$ was determined 
phenomenologically from the energy per particle of a unitary Fermi gas at 
$T=0$, given by Eq. (\ref{eq1}). A strongly interacting Fermi gas at
unitarity has no length scale other than the inverse of the Fermi
momentum. Consequently, its potential energy has the same
$k_F$-dependence as that of the kinetic energy. Since the potential is
attractive, the parameter $\xi$ is less than unity. 
The parameter $\xi=0.44$ is close to its experimental value~\cite{carlson} 
and therefore the statistical parameter is $g=\xi^{3/2}=0.29$. Since the 
value of the statistical parameter is dependent only on the nature of the 
interaction, it is fixed once and for all at all temperatures. Using this 
value of $g$ and the distribution for FES particles given by
Wu~\cite{wu}, the average energy as a function of temperature was 
calculated. The agreement with the Quantum Monte-Carlo calculation 
(QMC)~\cite{bulag,burovski} for a homogeneous gas was found to be 
satisfactory. The idea was then extended to harmonically trapped gases 
where it was found to agree not only with earlier
calculations~\cite{hu}, but also with the available experimental 
data~\cite{kinast} (See Fig. 2, which is taken from ~\cite{bhaduri}). 
The approach in our earlier work  
was phenomenological, involving just one scale independent parameter $g$. 
In FES, it has been shown~\cite{shankar} that $g$ is determined by the 
high-temperature limit of the second virial coefficient. Our purpose in 
this paper is to determine the statistical parameter $g$ from the
second virial coefficient. The second virial coefficient is obtained
from the inverse square potential in terms of $g$   
using a semiclassical procedure, which is known to reproduce the exact 
quantum result~\cite{langer}. Liu {\it et. al} have also obtained it 
directly from the quantum spectrum given in Fig. (\ref{fig1}).  
Equating these two second virial coefficients, 
we find that $g=1-1/\sqrt{2}\simeq 0.29$. With this result, our FES
calculations done earlier~\cite{bhaduri} require no free parameters
any more. 
\begin{figure}[htp]
\centering
\includegraphics[width=7cm]{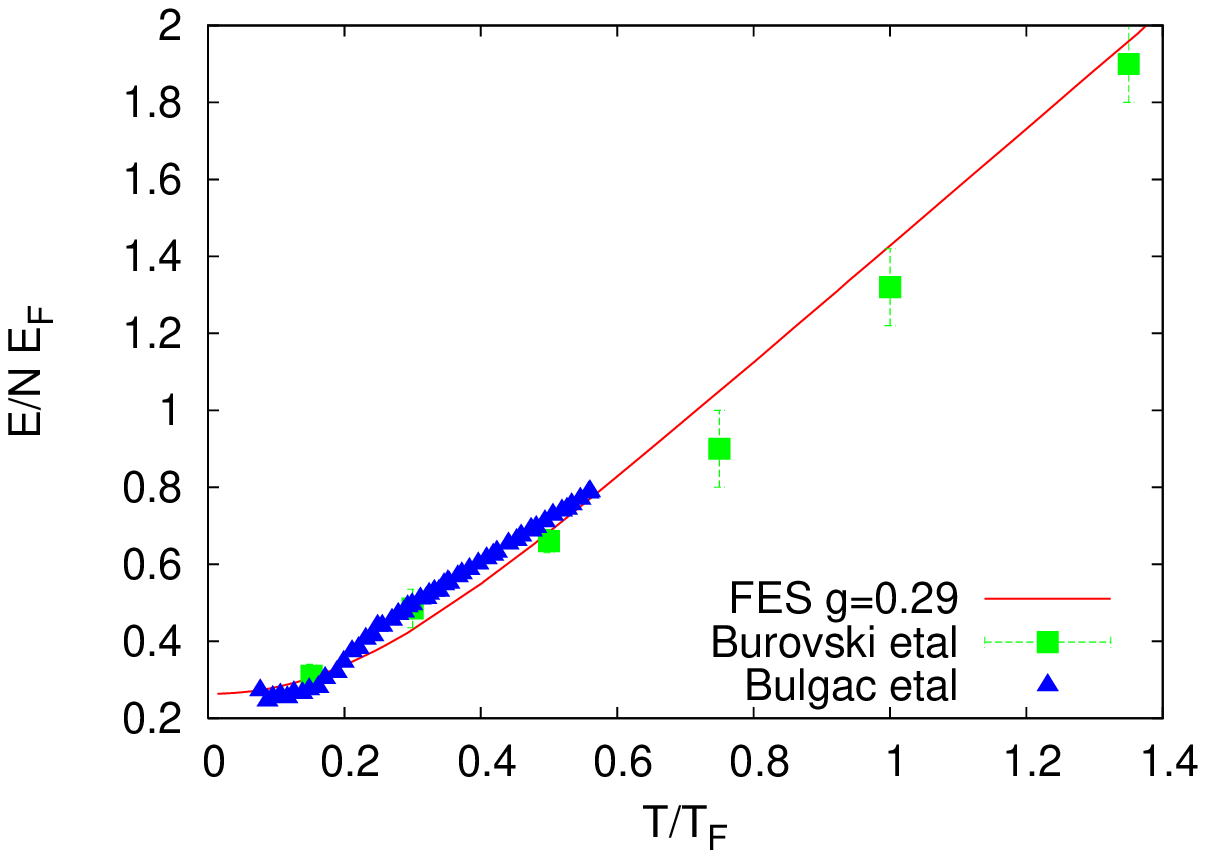}
\includegraphics[width=7cm]{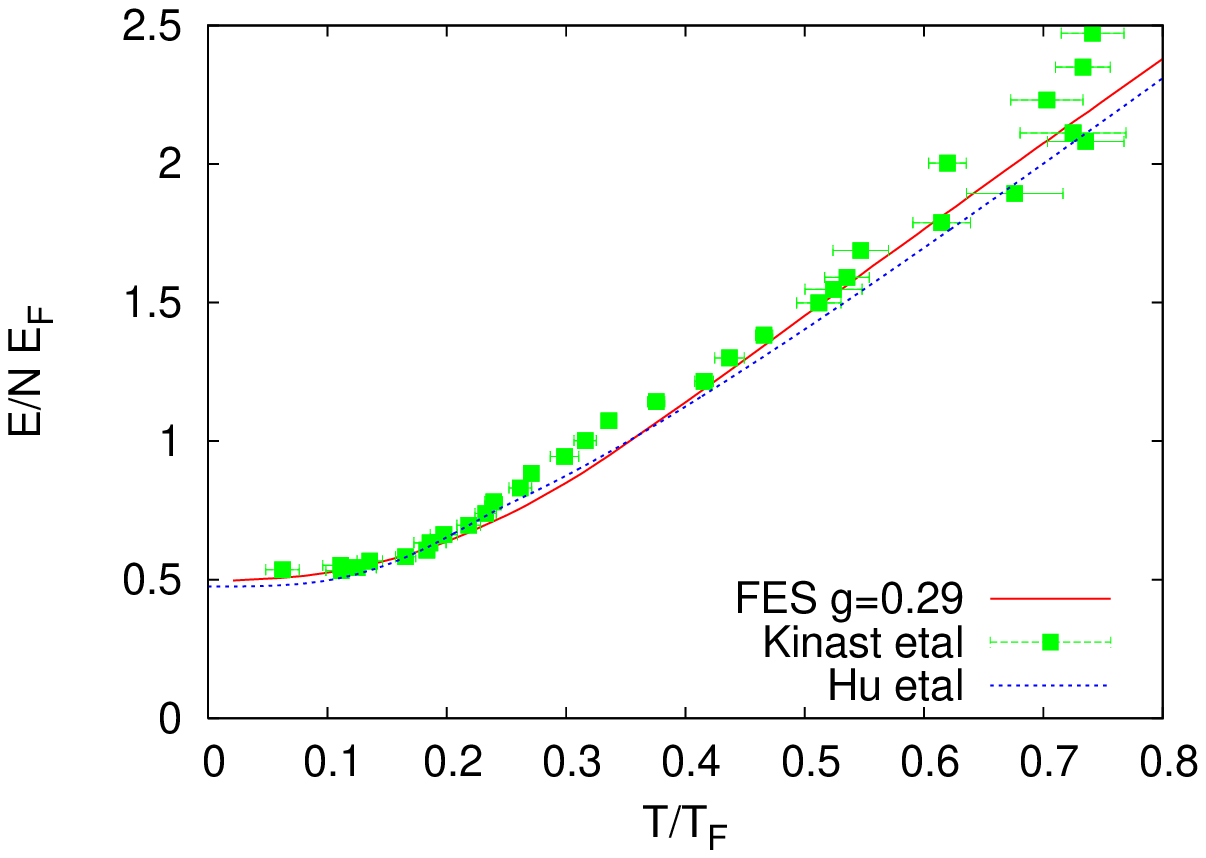}
\caption{Energy per particle as a function of temperature. At left we 
give the results for a homogeneous gas (solid line). Our results are 
compared with the MC calculations of ref.\cite{burovski} (solid squares) 
and ref.\cite{bulag} (solid triangles). On the right the results for a 
harmonically confined system is shown (solid line). The dashed line 
corresponds to the calculations presented in ref.\cite{hu} and the 
experimental data are from ref.\cite{kinast}. See \cite{bhaduri} for more 
details.}
\label{fig2}
\end{figure}

The relative s-wave two-body spectrum of Fig.(\ref{fig1}) was calculated 
for two interacting atoms in a spherical harmonic oscillator by Busch et. 
al~\cite{busch}. The interaction is of zero range, and its scattering 
length is tuned to infinity. The spectrum is universal, and is practically 
unaltered for interaction potentials whose range is much shorter than the 
oscillator length. In section 2, we briefly recapitulate the essentials of 
this spectrum. In section 3, we show that this spectrum is also reproduced 
by an equivalent one-dimensional system, namely the Calogero-Sutherland 
model (CSM)~\cite{csm}, with an inverse square interaction. This lends 
credence to the connection of the unitary spectrum to FES, since it is 
well known that CSM provides an exact realisation of FES. 
Section 4 contains the main theme of the paper. Here $g$ is obtained
from the high-temperature second virial coefficient of the gas, as
outlined earlier in this section. In FES, once the temperature
parameter $g$ is fixed, the properties of the gas are determined at all 
temperatures. 
Our calculations show that not only is the cold-atom two-body problem at 
$T=0$ mimicked by the two-body Calogero model, but the same applies
for the many-body case at finite $T$.

\section{Two-body spectrum at unitarity }

Consider two identical fermionic atoms in different spin states 
interacting with a zero-range potential. When the strength of the 
interaction is tuned to produce a zero-energy bound state, the scattering 
length $a=\pm \infty$. When these two atoms are trapped in a 
three-dimensional spherical harmonic oscillator, the full spectrum is 
analytically calculated and is well-known~\cite{busch, jonsell, vanzyl}. 
The s-wave relative spectra, after subtracting the Centre-of-Mass (CM) 
energy, for the lowest few states are shown in Fig.(\ref{fig1}). It is 
important to note that the only effect of the interaction is to produce a 
constant unit downward shift in the energy levels, which are labelled in 
units of the oscillator spacing $\hbar\omega$. As pointed out earlier, 
this behaviour is also a characteristic of CSM~\cite{csm} with an inverse 
square interaction. The un-normalised ground state eigenfunctions in 
relative coordinates, $u_0(r)=r\psi_0(r)$, for the two cases are given by
\bea
u_0(r)&=&r\exp(-r^2/2),~\mbox{non-interacting},\nonumber\\
u_0(r)&=&\exp(-r^2/2), ~~~\mbox{interacting}, ~~~~~~~r>0~.
\label{eq2}
\eea
where the relative distance $r$ is dimensionless, expressed in units of
the oscillator length $L=\sqrt{(\hbar/m\omega)}$, and $m=M/2$, $M$ being
the mass of the atom. The tower of states built on the ground state are
the nodal excitations. Note that in the interacting case the wave function
actually corresponds to the irregular solution of the non-interacting
system that is normally excluded as an eigenstate. However, this is a 
valid
solution at unitarity due to the presence of the singular interaction at
the origin~\cite{vanzyl}. The spectra shown above remains almost 
unchanged even when the interaction range is finite, provided the
latter is much smaller than the oscillator length~\cite{vanzyl}.

In the next section we illustrate a one dimensional template wherein such
a spectrum is realised.

\section{s-wave spectrum from Calogero model}

We now demonstrate that the two-body s-wave spectra shown above may
be generated by the one dimensional Calogero Hamiltonian~\cite{csm}. We  
start with the two-body Hamiltonian
\beq
H=\frac{(p_1^2+p_2^2)}{2M} +\frac{1}{2} M\omega^2 (x_1^2+x_2^2) +
\frac{\hbar^2}{M} \frac{g(g-1)}{x^2}`,
\label{ham}
\eeq
where the interaction strength is controlled by $g\geq 0$. We may
transform to CM co-ordinates 
$$P=(p_1+p_2), ~X=(x_1+x_2)/2;$$
and relative co-ordinates 
$$p=(p_1-p_2)/2, ~x=(x_1-x_2).$$
The CM Hamiltonian is that of a single oscillator and does not play any 
further role. Consider the relative Hamiltonian given by 

\beq
-\frac{\hbar^2}{M}\frac{d^2}{dx^2}\psi(r)+\left(\frac{1}{4}M\omega^2 
x^2+\frac{\hbar^2}{M}\frac{g(g-1)}{x^2}\right)\psi(x)
=E \psi(x).
\label{csm}
\eeq
The particles cannot cross in one dimension and we may restrict the
solutions to the region $0\le x \leq \infty$ (this mimics the range of 
$r$ in three dimensions). Because of the singular
nature of the interaction, the solutions go to zero at the coincident
point. We define $r=x/L$, and express the energy in units of $\hbar\omega$. 
The physically acceptable ground state solution in the interval
$0< r <\infty$ and its energy are again given by
\bea
\psi_0(r)&=&r^g \exp(-r^2/2), \nonumber\\
E_0&=&(g+1/2)~.
\label{zero}
\eea

The full spectrum of states is easily found, with the energy eigenvalues
and the corresponding eigenstates given by
\bea
E_n&=&(2n+g+1/2)~, \hspace{2cm} n=0,1,2,3,..\\
\psi_n(r)&=&r^g \exp(-r^2/2) L_n^{g-1/2}(r^2)~.
\label{spec}
\eea
Furthermore, following Calogero, the physically acceptable solutions
may be extended to the whole range $-\infty < x <\infty $ by imposing the 
condition
\beq
\psi(-x)=\pm \psi(x)~.
\label{sign}
\eeq
From the above symmetry/antisymmetry condition, we note that $g=1$ 
corresponds to noninteracting ``fermions", and $g=0$ to ``bosons".
This interpretation in one dimension should not be taken literally, 
since the permutation of the particles by crossing is not allowed.  
Comparing Eqs.(\ref{eq2}) and (\ref{zero}), the spectra (a) and (b) in 
Fig.1, we see that (a) corresponds to $g=1$, and (b) to $g=0$. 
It is 
important to note that the interaction vanishes at both $g=1$ and $g=0$, 
but the tower of energy states are not identical. In the first case, 
$g=1$, the states correspond to the non-interacting system shown in the 
s-wave spectra of the three dimensional system in Fig.(\ref{fig1}). The 
interaction is attractive for $g<1$ and repulsive for $g>1$. The maximum 
attraction is precisely at $g=1/2$.  In the second case, $g=0$, is 
approached from the attractive side and the spectrum is identical to the 
interacting case shown in Fig.(\ref{fig1}).

It has been shown by several authors~\cite{murthy} that the 
interacting particles of CSM may be regarded as non-interacting 
quasi-particles obeying FES with $g$ as the statistical parameter as 
defined by Haldane~\cite{haldane}. In this sense the spectra shown in 
Fig.(\ref{fig1}) are remarkable. In the light of CSM we may physically 
interpret the s-wave spectrum obtained using the pseudo-potential as 
similar to the phenomenon in which the interaction is statistical in 
nature which produces the effect of turning fermions into bosons. While 
this analogy is indicative of the nature of the interaction as 
statistical, the actual value of the statistical parameter for the which 
the spectra agree, namely $g=0$ in one dimension, cannot be interpreted 
literally in three dimensions. 

\section{ The universal second virial coefficient}

The grand partition function of a system may be expanded as a series in 
fugacity parameter, $z=\exp(-\beta\mu)$, at high temperatures (or low 
densities). The second virial coefficient $a_2=-b_2$, where $b_2$ is the 
so-called cluster integral~\cite{pathria}. The second virial coefficient, 
to a large extent, determines the thermodynamic properties of a dilute 
interacting gas. If in particular the interaction is statistical in the 
sense defined by Haldane, the second virial coefficient is related to the 
statistics parameter $g$ in the high temperature limit and plays an 
important role in determining systems which obey FES~\cite{shankar}. A 
classic example is the Calogero model where a gas with inverse-square 
pairwise interaction can be regarded as an ideal gas obeying 
FES~\cite{murthy}.

The contribution to the interacting part of the second cluster
integral may be expressed in terms of the spectra of interacting and
non-interacting two particle systems and is given by
\beq
\Delta b_2={b}_2-{b}_2^0=\Sigma_n (\exp{-(\beta
E_n)}-\exp(-\beta E_n^0)),
\label{db2}
\eeq
where $E_n~(E_n^0)$ corresponds to the relative energy of the interacting
(non-interacting) system.
Substituting for the spectra as shown in Fig. 1, we then get
\bea
\Delta b_2 &=&\left(\frac{\exp(-y/2)}{1-\exp(-2y)}-
\frac{\exp(-3y/2)}{1-\exp(-2y)}\right)\\
         &=&\frac{\exp(-y/2)}{1+\exp(-y)}~,
\label{row}
\eea
where $y=\hbar\omega\beta$.
Taking the high-temperature limit $y\rightarrow 0$, we get
\beq
\Delta b_2~=~\frac{1}{2}-\frac{y^2}{16} +...
\label{one}
\eeq
Although calculated in harmonic confinement, the temperature-independent 
value of $1/2$ is universal, and is also valid in a homogeneous gas. This 
is in agreement with the result of Beth and Uhlenbeck~\cite{beth}, and
Ho and Mueller~\cite{ho} who obtained the universal value of $1/2$ at 
resonance for a homogeneous gas.

In the above, we have not included spin factors. If the spin factors are 
included in the one-body canonical partition function we have $\Delta b_2 
= 1/4$ in agreement with the pseudo-potential calculation of Liu et 
al~\cite{liu}. 
The overall factor of $1/2$ due to spin will be omitted also in the 
subsequent semiclassical calculation to be consistent. We exploit this 
universal value of $\Delta b_2$ at the s-wave resonance  
to provide a microscopic explanation of 
the origin of FES in cold fermionic atoms. The CSM type inverse-square 
interaction is scale invariant and gives rise to a 
temperature-independent 
universal second virial coefficient in any dimension. 
We therefore use the semiclassical approach to provide a link between 
$\Delta b_2$ and strength of inverse square interaction. In the partial 
wave decomposition we have
\beq
\Delta b_2=\sum_{l=0}^{l=\infty}(2l+1)\Delta b_2^{(l)},
\label{db22}
\eeq
where the contribution due to the interacting part, that is $\Delta
b_2^{(l)}$,
can be written in the semiclassical WKB approximation as
\beq
\Delta b_2^{(l)} = \frac{1}{\lambda}\int_0^\infty dr~\exp\left[
-\beta\frac{\hbar^2l(l+1)}{Mr^2}\right][\exp(-\beta V(r))-1],
\label{db23}
\eeq
where $\lambda=\sqrt{2\pi\hbar^2\beta/M}$ is the thermal wave length,
$V(r)$ is the two-body potential. Furthermore
summing over all the partial waves, treating $l$ as a continuous variable,
we get
\beq
\Delta b_2~=~\frac{2\pi}{\lambda^3}\int_0^\infty r^2~dr~[\exp(-\beta
V(r))-1],
\label{db24}
\eeq
which is indeed the correct semiclassical expression~\cite{pathria}. 
In general, the classical $\Delta b_2^{(l)}$ in Eq.(\ref{db23}) even at 
resonance depends on temperature as 
well as the parameters of the potential. The lowest order WKB 
approximation is poor at resonance, and the universality is lost. The only 
exception to this rule is the scale invariant inverse square potential. 
More over, when the Langer modification~\cite{langer} of replacing 
$l(l+1)$ by $(l+1/2)^2$ is implemented, the WKB approximation reproduces 
the quantum results exactly. As seen from the one dimensional example, the 
s-wave asymptotic wave function is exactly reproduced at resonance when 
the scattering length $a \rightarrow \infty$. Note however that the 
inverse square potential is {\it only} applicable in the $l=0$ partial wave at 
resonance. The inverse square potential is a long-range potential, and
one may ask why its contribution to the second virial coefficient  
from the $l>0$ channels are not included. To answer this, recall that we are
calculating the {\it interaction} part of the second virial
coefficient. Due to the FR in the $l=0$ partial wave, the interaction
is dominant only in this channel, and the higher partial waves
contribute negligibly. In Busch et al.'s paper~\cite{busch}, 
the two-body spectrum is is obtained from a pseudopotential that acts only 
in the s-state. This spectrum is also used by Liu {\it et. al}~\cite{liu}.
At unitarity, there is no length scale due to the
interaction, and we take the effective interaction to be inverse
square, only applicable in the s-state. Away from the unitary point, 
the fractional exclusion statistics (FES) is not applicable. 

We therefore assume that the two body s-wave  
potential in relative coordinates is given by
\beq
V_0(r) = \frac{\hbar^2}{M} \frac{g(g-1)}{r^2}.
\eeq
Substituting this in Eq.(\ref{db23}), setting $l=0$, and implementing
the Langer correction, we get
\bea
\Delta b_2^{(0)} &=& \frac{1}{\lambda}\int_0^\infty dr~\exp\left[
-\beta\frac{\hbar^2}{4Mr^2}\right][\exp(-\beta
V_0(r))-1]\label{kyabat}\\
&=&\frac{1}{\sqrt{2}}[\frac{1}{2}-\sqrt{(g-1/2)^2}].
\label{db25}
\eea
Equating this to the universal value of $\Delta b_2$ obtained from the 
s-wave spectrum in Eq.(\ref{one}), we have
\beq
\Delta b_2
~=~\frac{1}{2}~=~\frac{1}{\sqrt{2}}[\frac{1}{2}\mp (g-\frac{1}{2})].
\label{db26}
\eeq
We have two solutions corresponding to the $g=1-1/\sqrt{2}\simeq 0.29$
and $g=1/\sqrt{2}=\approx 0.71$. The solution $g=0.29$ is appropriate 
for an attractive interaction in the fermionic basis. Note that this 
result would not change 
by taking the spin into account, as both sides of Eq.(\ref{db26})
would be multiplied by $1/2$. 
An identical result is obtained when
instead of a homogeneous interacting gas, we put the particles in an
oscillator trap. Now Eq.(\ref{kyabat}) is given by   
\beq
\Delta b_2^{(0)} = \frac{1}{\lambda}\int_0^\infty dr~\exp\left[
-\beta\frac{\hbar^2}{4Mr^2}\right][\exp(-\beta
V_0(r))-\exp(-\frac{\beta}{4}M\omega^2 r^2)]~,
\label{accha}
\eeq
where 
\beq
V_0(r) = \frac{1}{4}M\omega^2 r^2+\frac{\hbar^2}{M} \frac{g(g-1)}{r^2}.
\eeq
The interaction parameter$g$ may still be interpreted as the statistical
parameter of FES since the mapping between the inverse-square interaction in
one dimension (CSM) and FES is exact. The result of the integration in 
Eq.(\ref{accha}) is 
\beq
\Delta
b_2^{(0)}=\frac{1}{\sqrt{2}\hbar\omega\beta}\left[\exp(-\hbar\omega\beta
\sqrt{(g-1/2)^2})-\exp(-\hbar\omega\beta/2)\right]
\eeq 
In the limit of $\hbar\omega\beta\rightarrow0$ the result is the same as 
in Eq.(\ref{db25}). It may be noted that the
harmonic oscillator merely acts as a regulator in the high temperature
limit, and yields the homogeneous gas result. 

The relation between the second virial coefficient and the statistics
parameter $g$ for a homogeneous gas in FES\cite{shankar} is given by
\beq
\frac{1}{2}-g = 2^{d/2}b_2,
\eeq
where d is the dimension of the space that is relevant. For s-wave 
contribution alone, we choose effectively $d=1$. We 
therefore have
\beq
1-g = \sqrt{2}\Delta b_2,
\label{low}
\eeq
where the non-interacting limit has $g=1$ for fermionic atoms. This
reproduces the solution given above for the interaction parameter $g$ in
CSM thus establishing the connection with statistics parameter of FES.

The ground state energy of a gas of FES particles is given by 
Eq.(\ref{eq1}). Even though this relation is for a three dimensional 
gas, we use $g$ obtained from the one-dimensional relation (\ref{low}) 
to determine $\xi$. This is because the deviation of the parameter $\xi$ 
from the noninteracting value of unity is due to the attractive 
potential energy. This potential energy arises from the pair-wise 
interaction that (as already noted) acts only in the $l=0$ state. So far 
as the potential energy is concerned, only a single partial wave is 
relevant, and the system is effectively one-dimensional.

We have seen that the value of $g$ obtained above from the 
high-temperature regime is compatible with the experimental results at 
all temperatures, thus establishing the connection between an ideal gas 
obeying FES and a dilute gas of strongly interacting fermionic atoms at 
unitarity.  The earlier phenomenological analysis of the average energy 
and the chemical potential at finite temperature using the distribution 
function for FES particles is also therefore justified. As further 
evidence, we may add the following observation, as detailed in the 
recent article by Bloch {\it et. al} ~\cite{dimer}. They comment on the 
pressure-energy relation $P=2E/3$ obeyed by a gas at unitarity. From 
this, it may be deduced that the effective two-body interaction 
potential has to be of inverse-square nature, given that it is not 
noninteracting. This also gives rise to the virial theorem given by Eq. 
(134) of their paper, which is in agreement with the experimental 
results of Thomas {\it et. al.}~\cite{tom}.

\ack We thank R. Shankar for bringing to our attention the paper by Liu et 
al \cite{liu}. RKB has profited from earlier conversations with Duncan
O'Dell. This research was supported by NSERC of Canada.

\end{document}